# Harnessing Vacuum Fluctuations to Shape Electronic and Photonic Behavior

Qing-Dong Jiang[1,2,3 *]

[1]Tsung-Dao Lee Institute, Shanghai Jiao Tong University, Shanghai, 201210, China
[2]Shanghai Branch, Hefei National Laboratory, Shanghai 201315, China
[3]Shanghai Research Center for Quantum Sciences, Shanghai 201315, China
[*]Email: qingdong.jiang@sjtu.edu.cn

Abstract:
Vacuum quantum fluctuations are an inescapable and fundamental feature of modern physics. By integrating cavity-enhanced or surface-modified vacuum quantum fluctuations with low-dimensional materials, a new paradigm—*vacuumronics*—emerges, enabling unprecedented control over both material properties and photonic responses at the micro- and nanoscale. This synergy opens novel pathways for engineering quantum light-matter interactions, advancing applications in quantum photonics, nanoscale optoelectronics, and quantum material design.

Main Text:
It may seem surprising that geckos climbing walls and Hawking radiation from black holes share the same underlying cause: vacuum quantum fluctuations. Research has shown that geckos adhere to surfaces via van der Waals forces—weak intermolecular attractions arising from transient electric dipoles induced by vacuum fluctuations [1]. On the other hand, Hawking radiation occurs when quantum fluctuations near a black hole's event horizon spontaneously produce particle-antiparticle pairs; when one particle escapes while its counterpart falls inward, the black hole gradually loses mass [2]. These ubiquitous fluctuations give rise to a wide range of fundamental phenomena, including the Casimir force, Lamb shift, Schwinger pair production, anomalous magnetic moments, and spontaneous emission—demonstrating their profound influence across physical scales from the atomic to the cosmological [3].

What is less widely recognized, however, is the potential of vacuum fluctuations to shape electronic behavior at the mesoscopic scale. Long regarded as a passive background to quantum phenomena, the vacuum is now being exploited as an active resource. When structured, confined, or amplified—through nanostructures, engineered surfaces, or optical cavities—vacuum fluctuations can be harnessed to modify, enhance, and control the electronic and photonic responses of materials. This emerging vision defines *vacuumronics*—a new frontier that treats vacuum fluctuations as a functional tool for next-generation electronics.

Quantum fluctuations can renormalize fundamental particle properties such as charge, mass, and energy. For example, a particle with a finite bare charge may appear completely screened by transient electron–positron pairs arising from vacuum fluctuations (Figure 1). To properly describe this phenomenon, one must invoke the renormalization group framework, which focuses solely on physically measurable quantities, while treating the bare particle as possessing an infinite, unobservable charge [3]. This idea carries broader implications: just as vacuum fluctuations renormalize the intrinsic properties of elementary particles, they can likewise alter the effective characteristics of quasiparticles in condensed matter systems—such as effective charge, effective mass, and band structure. This principle lies at the core of vacuumronics, which seeks to harness quantum vacuum fluctuations as a powerful tool for engineering and controlling material properties.

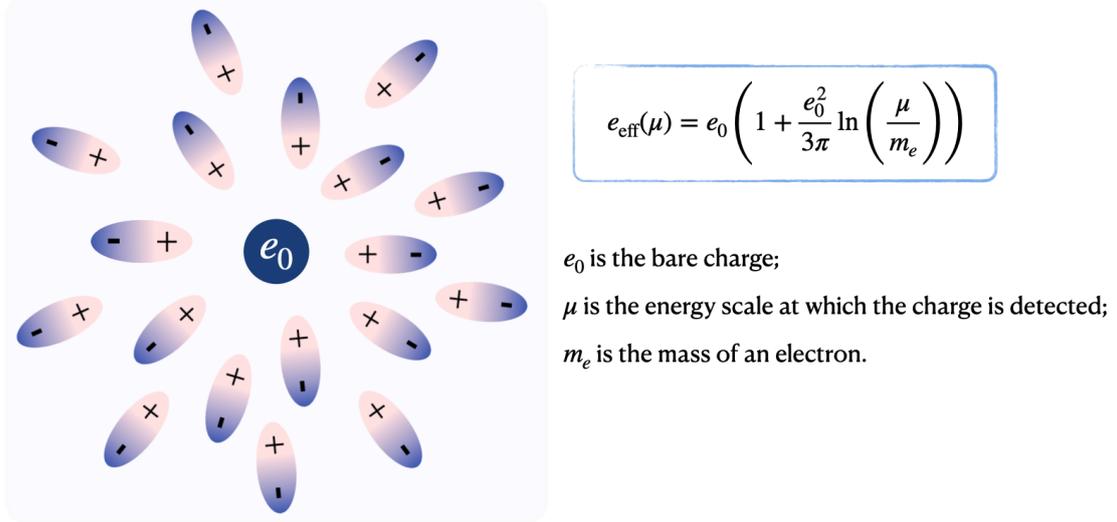

$$e_{\text{eff}}(\mu) = e_0 \left( 1 + \frac{e_0^2}{3\pi} \ln\left(\frac{\mu}{m_e}\right) \right)$$

$e_0$ is the bare charge;

$\mu$ is the energy scale at which the charge is detected;

$m_e$ is the mass of an electron.

Figure 1: Quantum fluctuations in the vacuum can generate transient electron-positron pairs which will be polarized by the bare charge $e_0$ and eventually completely screen the original bare charge at small energy scales (i.e. at large distances).

While quantum fluctuations are widely acknowledged for their foundational role in renormalizing physical properties, their profound impact on symmetry principles—the cornerstone of modern physics—remains underappreciated. In fact, quantum fluctuations influence system symmetries in two fundamental ways. First, they can break continuous symmetries that are preserved at the classical level, leading to quantum anomalies [4]. According to Noether's theorem, each continuous symmetry corresponds to a conserved current; quantum fluctuations break the symmetry, rendering formerly conserved currents non-conserved (Figure 2a). Second, quantum fluctuations can transmit broken symmetries from a material into the surrounding vacuum, creating an extended spatial zone - termed the quantum atmosphere - in the vicinity of the material. Within this quantum atmosphere, nearby quantum systems can be influenced or manipulated without direct contact. For example, a topological material known as a Chern insulator, which breaks time-reversal symmetry, can induce an anomalous Zeeman splitting for a spin placed within its quantum atmosphere (Figure 2b). This effect arises from a spin-dependent Lamb shift, induced by modified vacuum fluctuations near the time-reversal symmetry breaking surface. Due to the massless nature of photons, the vacuum quantum fluctuation induced Zeeman splitting follows the power law scaling:

$$\Delta E^{\text{QF}} = \frac{(10 \text{ nm})^2}{r^2} \times 0.06 \text{ } \mu\text{eV}$$

where $r$ is the distance from the spin to the material's surface. At a distance of $10\,\text{nm}$, this corresponds to a splitting of approximately $0.06\,\mu\text{eV}$—comparable to the effect of a $10\,\text{Gauss}$ magnetic field. Importantly, the quantum atmosphere effect decays as a power law $\sim 1/r^2$ rather than exponentially, allowing the quantum atmosphere to extend over distances of $100\,\text{nm}$ or more. Moreover, the divergence of the effect as $r \to 0$ sharply contrasts with classical stray magnetic fields, which typically saturate at short distances. This distinction provides a clear experimental signature to differentiate between classical field-induced effect and those arising from the quantum atmosphere. These remarkable characteristics suggest that the symmetry-breaking properties of a material can be both detected and utilized through their imprint on the surrounding vacuum—the quantum atmosphere—opening new avenues for non-contact control and tailoring of nearby quantum systems.

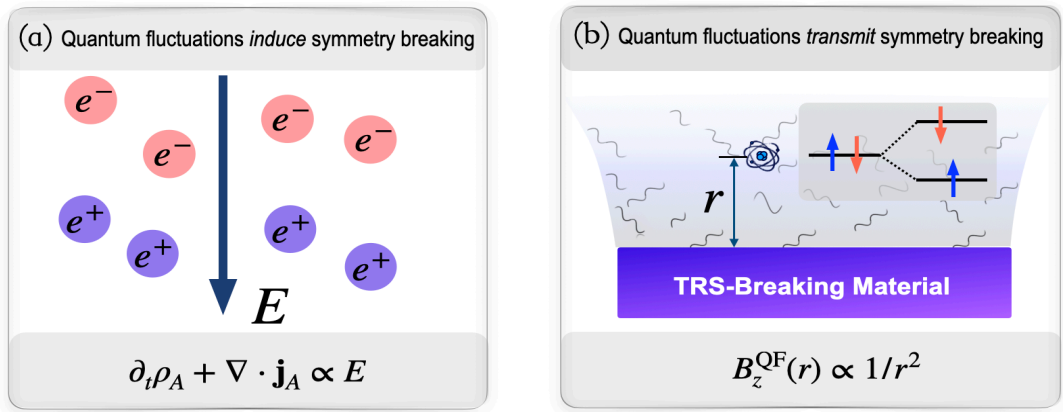

Figure 2: (a) In a one-dimensional massless Dirac system, quantum fluctuations break the so-called axial symmetry, resulting in a non-conserved axial current $\mathbf{j}_A$—defined as the difference between the positive and negative charge currents. Under an external field, this results in the separation of positive and negative charges—a phenomenon also known as the Schwinger effect. (b) Vacuum quantum fluctuations can transmit a material's symmetry breaking into the surrounding vacuum. Within the quantum atmosphere of a Chern insulator that breaks time-reversal symmetry, a nearby spin experiences an effective magnetic field that decays as a power law with the distance $r$ from the material's surface.

A cornerstone of quantum optics is the confinement of light within high-finesse optical cavities, which—by drastically reducing the modal volume—can amplify quantum fluctuations to unprecedented levels, enabling extreme light-matter coupling regimes [6]. For instance, a cavity with an average mode volume of 1 nm³ can induce enormous fluctuating electric fields with strengths reaching the order of $10^9$ V/m. This foundational insight has recently converged with advances in two-dimensional materials, allowing the transfer of cavity quantum fluctuation effects from traditional quantum optical systems to condensed matter platforms [7]. Remarkably, these vacuum electromagnetic fluctuations—often referred to as virtual photons—can now manipulate fundamental material properties, including band structure, topological states, magnetic ordering, and even superconductivity—giving birth to the emerging field of cavity materials engineering [8-12]. The unique advantage of this approach lies in its field-free control mechanism: cavity vacuum fluctuations modify material properties without requiring static or dynamic real fields. This circumvents the limitations of conventional field-based techniques like Floquet engineering, which often introduce undesirable heating, quantum decoherence, and challenging-to-control non-equilibrium states. By harnessing the quantum nature of vacuum fluctuations, cavity quantum engineering opens a new pathway for manipulating quantum states of matter.

The integration of symmetry-breaking effects with cavity quantum electrodynamics has opened new avenues for controlling material properties, with symmetry-breaking cavities offering particularly broad tunability. Remarkably, even in symmetry-preserving cavities, spontaneous symmetry breaking can emerge and induce interesting quantum phenomena such as emergent gyrotropic Hall effect [13]. While real-field-driven symmetry breaking produces fundamental effects like the quantum Hall effect and Zeeman splitting, researchers have shown that quantum-fluctuation in a symmetry-breaking cavity or near a symmetry-breaking material surface can similarly induce these phenomena (Fig. 3) [5,8,9]. In these settings, light–matter hybridization occurs without real photons, allowing virtual photons to renormalize material properties in equilibrium. This fluctuation-driven mechanism offers a pristine platform for quantum control, free from the heating and dynamical instabilities that typically accompany externally driven (e.g., Floquet) systems, and thus circumventing the limitations of traditional field-driven methods.

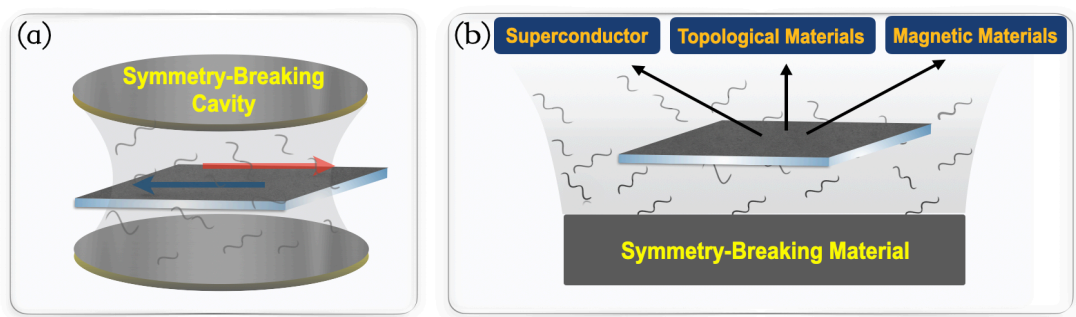

Figure 3: (a) Cavities can imprint their symmetry-breaking features onto quantum-fluctuation-induced phenomena, providing a novel tuning mechanism for engineering quantum materials—without the need for static fields or optical illumination. (b) Quantum fluctuations transmit the symmetry breaking of a material to its quantum atmosphere, potentially useful for engineering quantum materials, such as superconductors, topological materials or quantum magnetic materials.

From a grander point of view, the quantum vacuum and quantum materials constitute a profoundly interconnected system in which vacuum fluctuations influence material properties while material responses reciprocally modify vacuum properties. This mutual interaction not only provides a neat approach to engineer quantum many-body phenomena but also enables transformative possibilities—exotic quantum phases in materials, particularly in strongly correlated systems, can dramatically alter optical phenomena. For instance, materials with tailored electronic correlations may enhance vacuum squeezing effects, establishing new paradigms in quantum optics.

Looking ahead, such cross-domain control unveils a new frontier where quantum materials act as active designers of photonic quantum devices. Through vacuum quantum fluctuations, materials such as topological insulators may imprint their intrinsic robustness and symmetry-protected properties onto surrounding electromagnetic fields, transferring their stability into the photonic domain. This capability paves the way for next-generation technologies—from topological lasers [14] and ultra-stable atomic clocks [15] to noise-resilient quantum computers [16]—where vacuum fluctuations are no longer mere background noise, but a tunable and valuable quantum resource.

**Acknowledgement:** I am grateful to useful discussions with Hong Ding at Tsung-Dao Lee Institute. This work was supported by National Natural Science Foundation of China (NSFC) under Grant No. 12374332, the Innovation Program for Quantum Science and Technology Grant No. 2021ZD0301900, Cultivation Project of Shanghai Research Center for Quantum Sciences Grant No. LZPY2024, and Shanghai Science and Technology Innovation Action Plan Grant No. 24LZ1400800.

**Competing Interests:** There is no competing interest of this work.